\documentclass[useAMS,usenatbib,fleqn]{mn2e}
\usepackage{epsfig}
\usepackage{amsmath}
\usepackage{color} 
\usepackage{natbib}
\usepackage{rotating}
\usepackage{graphicx}
\usepackage[colorlinks=true,citecolor=blue]{hyperref}
\def\ew{EW}

\def\erg{~$\rm ergs~s^{-1}~$}

\def\civ{C~{\sc iv}}
 
\def\siv{Si~{\sc iv}}

\def\nv{N~{\sc v}}

\def\zabs{$z_{\rm abs}$} 
\def\zem{$z_{\rm em}$}

\def\chinu{$\chi^{2}_{\nu}$} 
\def\chisq{$\chi^{2}$} 
\def\kms{km s$^{-1}$} 
\def\cmss{cm s$^{-2}$}
\def\kmsy{km s$^{-1}$ yr$^{-1}$}
\title[\civ\ BAL variability]{ 
\civ\ absorption line variability in X-ray bright BALQSOs}
 \author[R. Joshi, H. Chand, R. Srianand and J. Majumdar]{Ravi Joshi $^{1}$\thanks{E-mail:
     ravi@aries.res.in (RJ); hum@aries.res.in (HC); anand@iucaa.ernet.in
     (RS);  jhilik3837@gmail.com (JM)},
   Hum Chand $^{1}$, Raghunathan Srianand $^2$, Jhilik Majumdar $^3$
   \\\\ $^{1}$Aryabhatta Research Institute of
   Observational Sciences (ARIES), Manora Peak, Nainital $-$ 263 002,
   India\\
   $^{2}$Inter-University Centre for Astronomy and Astrophysics (IUCAA), Postbag 4, Ganeshkhind, Pune 411 007, India \\
   $^{3}$Jadavpur University, Kolkata 700 032, India}

\begin{document}
\date{Accepted ---. Received ---; in original form ---}

\pagerange{\pageref{firstpage}--\pageref{lastpage}} \pubyear{2014}

\maketitle

\label{firstpage}
\begin{abstract}

We report kinematic shift and strength variability of \civ\ broad
absorption line (BAL) trough in two high-ionization X-ray bright QSOs
SDSS J085551$+$375752 (at \zem\ $\sim$ 1.936) and SDSS
J091127$+$055054 (at \zem\ $\sim$ 2.793). Both these QSOs have shown
combination of profile shift, appearance and disappearance of
absorption components belonging to a single BAL trough. The observed
average kinematic shift of whole BAL profile resulted in an average
deceleration of $\sim - 0.7 \pm 0.1$, $- 2.0 \pm 0.1$~\cmss\ over a
rest-frame time-span of 3.11 yr and 2.34 yr for SDSS J085551$+$375752
and SDSS J091127$+$055054, respectively. To our knowledge, these are
the largest kinematic shifts exceeding by factor of about 2.8, 7.8 than the
highest deceleration reported in the literature; making both of them
as a potential candidate to investigate outflows using
multi-wavelength monitoring for their line and continuum variability.
 We explore various possible mechanisms to understand the observed
  profile variations.  Outflow models involving many small
  self-shielded clouds moving probably in a curved path provides the
  simplest explanation for the \civ\ BAL strength and velocity
  variations along with the X-ray bright nature of these sources.
\end{abstract}
\begin{keywords}
galaxies: active - quasars: absorption lines - quasars: general quasars: individual:
J085551$+$375752, J091127$+$055054.
%% galaxies: distances and redshifts, quasars: absorption lines --
%% quasars:  general -- techniques: spectroscopic -- quasar : individual (J085551$+$375752, J091127$+$055054) 
\end{keywords}

%\maketitle

\section{Introduction}
\label{lab:xbal_intro}
About 10$-$20 per cent of the QSO population shows strong blueshifted
broad absorption lines (henceforth called BALQSOs), with velocity widths greater than
2000~\kms\ and typical outflow velocities of 1000$-$30,000~\kms
\citep[e.g.,][]{Weymann1991ApJ...373...23W}. This has been interpreted
as a signature of outflows from the accretion disc. Outflows play an
important role in controlling the growth of the central massive black
hole, the evolution of the host galaxy and the chemical enrichment of
the inter-galactic medium
\citep[e.g.,][]{Ostriker2010ApJ...722..642O}. However, the basic
physical conditions, acceleration mechanism(s), location and three
dimensional structure of QSO outflows are poorly understood. Line
variability study is one of the powerful methods which can provide
useful insights into the structure and dynamics of the
outflowing gas. In the case of BALQSOs such line variability is
commonly reported as changes in the absorption strength
\citep[e.g.][]{Hamann1997ApJ...478...87H,Srianand2001A&A...373..816S,Misawa2005ApJ...629..115M,
  Lundgren2007ApJ...656...73L, Hall2011MNRAS.411.2653H,
  Capellupo2012ASPC..460...88C,Capellupo2012MNRAS.422.3249C,
  Capellupo2013MNRAS.429.1872C, Ak2013ApJ...777..168F,
  Vivek2014MNRAS.440..799V}; and/or appearance, disappearance of
absorption trough \citep[e.g.,][]{Hamann2008MNRAS.391L..39H,
  Hidalgo2011MNRAS.411..247R,
  Vivek2012MNRAS.421L.107V,Vivek2012MNRAS.423.2879V,
  Hamann2013arXiv1302.0201H}. However, the kinematic shift in BAL
profiles have also been seen in very few cases
\citep[e.g.,][]{Vilkoviskij2001MNRAS.321....4V,
  Gabel2003ApJ...595..120G, Hall2007ApJ...665..174H}. The observed
behavior of appearance or disappearance of BAL trough, their
absorption strength variation and the kinematic shift in absorption
profile are most readily understood as a result of (i) changes in the
ionization state as a function of velocity in a fixed outflow; (ii)
changes in the acceleration profile and/or geometry of the outflow due
to change in the driving force or mass-loss rate; (iii) by actual line
of sight acceleration of a shell of material from a continual flow;
and, (iv) due to transverse motion of the absorbing cloud(s) relative
to the line of sight \citep{Gabel2003ApJ...595..120G,
  Hall2007ApJ...665..174H, Lundgren2007ApJ...656...73L,
  Capellupo2011MNRAS.413..908C, Vivek2012MNRAS.421L.107V}. \par

  In a general scenario, BAL outflows are believed to arise from 
    the QSO's accretion disc and driven by the radiation pressure
  \citep{Arav1994ApJ...427..700A,Murray1995ApJ...451..498M,Proga2000ApJ...543..686P,
    Proga2004ApJ...616..688P}. However, the required radiation to push
  the outflow at a high relativistic speed may also over-ionize
  the gas and hence make it transparent to the radiation that drives
  the flow.  This problem is resolved by proposing the radiative
  shield only to be close to the base of the outflow
  \citep{Murray1995ApJ...451..498M}. This has also been used to
  explain the X-ray weakness of BALQSOs compared to normal QSOs by
  attributing the X-ray weakness to the absorption due to  high
    {H~{\sc i}} column densities $\rm (N_{\sc H})$ in the range of
  $10^{22} - 10^{24}$~cm$^{-2}$, due to radiative shield closer to the
  disc plane
  ~\citep[e.g.,][]{Wang1999ApJ...519L..35W,Proga2004ApJ...616..688P,
    Gallagher2006ApJ...644..709G, Stalin2011MNRAS.413.1013S}. However,
  origin of the X-ray weakness in BALQSOs is a matter of debate until
  now.\par

  Further, the observations of narrow absorption line outflows ($\sim$
  2000~\kms, some times refer to as ``mini-BALs'') have extended the
  above canonical picture by assigning these mini-BAL outflows to the
  sight lines at higher latitudes that perhaps skim the edge of main
  BAL flows farther above the disc \citep{Ganguly2001ApJ...549..133G,
    Hamann2008MNRAS.391L..39H}. This was also supported by the
  observed weak X-ray absorption in mini-BALs, if the X-ray absorbers
  resides primarily near the accretion disc as proposed above.
  Additional complication arises when we consider that mini-BALs also
  possess the very high-speed and moderate ionization as normal BALs,
  even without the protection of a radiative shield
  \citep{Hamann2013arXiv1302.0201H}. This might suggest that the shield
  may not be a critical feature of the wind, and hence, perhaps the
  models involving continuous flow with radiative shield may need
  replacement with the models involving many small self shielded
  clouds with low volume filling factor and driven out by radiative
  force while being confined by magnetic pressure
  \citep{deKool1995ApJ...455..448D, Rees1987QJRAS..28..197R}. Such
  clouds with magnetic confinement are also shown to have super
  thermal velocity dispersion and therefore only few of them can
  explain the observed broad and smooth BAL profiles \citep[e.g.,
    see][]{Bottorff2000MNRAS.316..103B,Hamann2013arXiv1302.0201H}. In
  such mechanism one would expect mixing of line shift and line
  strength variability introduced by small multiple clouds in contrast
  to  a  smooth variability in models of homogeneous outflows. The
  observational signatures for such scenarios are still awaited, for
  which study of the BALQSOs having broad absorption trough (like normal
  BALs) as well as being X-ray bright in nature (like mini-BALs) will
  be ideal candidates. Interestingly, such new population of X-ray
  bright BALQSOs have been discovered in recent X-ray surveys
  \citep[e.g.,][]{Giustini2008A&A...491..425G,
    Gibson2009ApJ...692..758G, Streblyanska2010AIPC.1248..513S,
    Stalin2011MNRAS.413.1013S}. The line-variability study of these
  sources may provide some useful hints towards understanding of  the above
  issues of outflow kinematics including the key question of BALQSOs
  being overall X-ray weak (i.e.,  either intrinsic or absorbed). \par

   Recently, we have started a pilot project for the spectral
   variability monitoring of 10 high-ionization BALQSOs detected in
   X-rays from the compilation of
   ~\citet{Giustini2008A&A...491..425G},
   ~\citet{Gibson2009ApJ...692..758G} and
   ~\citet{Streblyanska2010AIPC.1248..513S}. This sample has been
   constructed with the BALQSOs having (i) optical to X-ray spectral
   index, {$\alpha_{ox}$\footnote{Ratio between the monochromatic
       luminosities L$_{\nu}$ at 2 keV and 2500~\AA\ as, $\alpha_{ox}
       \equiv
       0.3838\log\biggl(\frac{L_{\nu}(2~\rm{keV})}{L_{\nu}(2500~\mathring{\rm
           A})}\biggr) $}}, greater than -1.8; (ii) the SDSS $g_{mag}
   < 19.0$, to achieve a good signal to noise ratio with 2-m class
   telescopes, in a reasonable  exposure time; and (iii) the redshift
   range of approximately $1.6 < z < 3.6$, in order to cover the
   \civ\ BAL trough in spectral range of 3800 $-$ 6840~\AA\ found most
   suitable in our observational plan (see below). Among our full
   sample we found two interesting cases of \civ\ BAL trough
   variation, here we present a detailed analysis of these systems.
   The detail results based on our full sample will be presented
   elsewhere. \par

   This paper is organized as follows. Section~\ref{lab:xbal_ANA}
   describes our analysis  including observations, data reduction
     and spectral analysis. In Section~\ref{lab:xbal_RES}, we present
     the results of our analysis, followed by a discussion in Section~\ref{lab:xbal_DnC} and
     finally our conclusions in Section~\ref{lab:xbal_con}.

\section{Analysis}
\label{lab:xbal_ANA}
\subsection{Observation and Data Reduction}
\label{lab:xbal_obsand dataredu}

 The observations were carried out using the IFOSC mounted on the 2
 meter telescope in IUCAA Girawali Observatory (IGO). We have taken a
 long-slit spectra covering the wavelength range 3800$-$6840~{\AA}
 using
 Grism{\footnote{\url{http://www.iucaa.ernet.in/~itp/etc/ETC/help.html\#grism}}}
 \#7 of IFOSC with a resolution $R \sim$~1140, to cover the \civ\ and
 \siv\ lines, respectively. A slit width of either 1.$^{\prime
   \prime}$0 or 1.$^{\prime \prime}$5 is used. Typical seeing during
 our observations were around 1.$^{\prime \prime}$2 to 1.$^{\prime
   \prime}$4. The raw CCD frames were cleaned using standard {\sc
   iraf}{\footnote{\textsc{iraf} is distributed by the
     \textsc{National Optical Astronomy Observatories}, which are
     operated by the Association of Universities for Research in
     Astronomy, Inc., under cooperative agreement with the National
     Science Foundation.}} procedures. The Halogen flats were used for
 the flat fielding the frames. We then extracted the one dimensional
 spectrum from individual frames using the \textsc{iraf} task
 ``apall''. Wavelength calibration of the spectra was performed using
 Helium-Neon lamp. The spectrophotometric flux calibration was done
 using standard stars and assuming a mean extinction for the IGO site.
 In cases of multiple exposures we co-added the flux with
 {${1}/{\sigma_i^2}$} weightage, where ${\sigma_i}$ is the error on
 the individual pixel. The error spectrum was computed taking into
 account proper error propagation during the combining process. The
 spectrum was corrected to vacuum helio-centric frame. \par

 The reduced one-dimensional spectra of BALQSOs for the comparison were
 downloaded from the SDSS and SDSS-BOSS Data Archive
 Server\footnote{\url{http://data.sdss3.org/bulkSpectra}}. Details about the
 SDSS spectral information can be found in
 ~\citet{York2000AJ....120.1579Y}. Briefly, these SDSS spectra cover
 a spectral range from 3800 to 9200~\AA, with a resolution
 ($\lambda/\Delta \lambda$) of about 2000 (i.e. 150~\kms). The BOSS
 spectra have wavelength coverage between $3600 - 10000$~\AA\ at a
 resolution of $1300 - 3000$ \citep[see][]{Dawson2013AJ....145...10D},
 as listed in column 5 of Table~\ref{lab:tab_sourceobs_info}.

\begin{table}
{\tiny
 \centering
 \begin{minipage}{130mm}
\caption{\scriptsize Log of observations and other basic parameters of the spectra.}
\label{lab:tab_sourceobs_info}
\begin{tabular}{@{}clc ccc cc@{}} 
\hline 
 \multicolumn{1}{c}{QSO} 
& \multicolumn{1}{l}{Instrument}  
& \multicolumn{1}{c}{Date} 
& \multicolumn{1}{c}{Exposure}
& \multicolumn{1}{c}{Resolution}
& \multicolumn{1}{c}{S/N \footnote{ Signal-to-noise ratio over the wavelength range 5800$-$6200 \AA.}}
\\
&
&\multicolumn{1}{c}{(MJD)}
&\multicolumn{1}{c}{Time (mins)}
&\multicolumn{1}{c}{(\kms)}
&
\\
 (1)&(2)&(3)&(4)&(5)&(6)
\\
\hline 
\\

                 &SDSS                                                                            & 52643  &55$\times$1    &150	 &16  \\
J0855$+$3757    &{IGO/IFOSC 7\footnote{ Wavelength coverage of 3800$-$6840~\AA.}}                & 55568  &45$\times$2  &310  &16 \\ 
                 &SDSS-BOSS                                                                       & 55973  &60$\times$1    &150	 &23  \\
 \hline
                &SDSS                                                                             & 52650  &52$\times$1    &150	 &24  \\
                &IGO/IFOSC 7	                                                                  & 55568  &45$\times$2    &310  &17 \\
                &SDSS-BOSS                                                                        & 55896  &60$\times$1    &150  &26 \\   
 J0911$+$0550   &IGO/IFOSC 7                                                                      & 55930  &45$\times$10\footnote{Among them 5 exposures belongs to observation taken after two \\ days, i.e., on MJD 55932 .}    &310  &48 \\   
                &IGO/IFOSC 1                                                                      & 55979  &45$\times$3    &370  &15 \\
\hline   
\end{tabular}                                                          
\end{minipage}    
}
\end{table}

\subsection{Continuum Fitting}

In order to have accurate measurements of variability in absorption
lines, one needs to carefully take into account the continuum as well
as the emission line  fluxes. This is necessary because of uncertainties
in the flux calibration and possible real changes in the QSO emission
lines. To model the continuum over the spectrum comprising of the
broad \civ\ absorption line region in the rest wavelength range
between $1270$ and $1800$~\AA, first we have used a single power law,
i.e. $a \lambda^{-\alpha}$, along with a lower (e.g., second) order
polynomial, constrained by the measured flux in emission and
absorption free regions namely $1323-1338$~\AA,
$1440-1450$~\AA\ (except for J085551$+$375752 having absorption) and
$1680-1800$~\AA\ in QSOs rest-frame. We use the emission redshift,
\zem, from ~\citet{Hewett2010MNRAS.405.2302H}, who have refined the
SDSS emission redshift values by reducing the net systematic errors by
almost a factor of 20, attaining an accuracy of up to ∼30~\kms. \par

\begin{figure*} 
\centering
\epsfig{figure=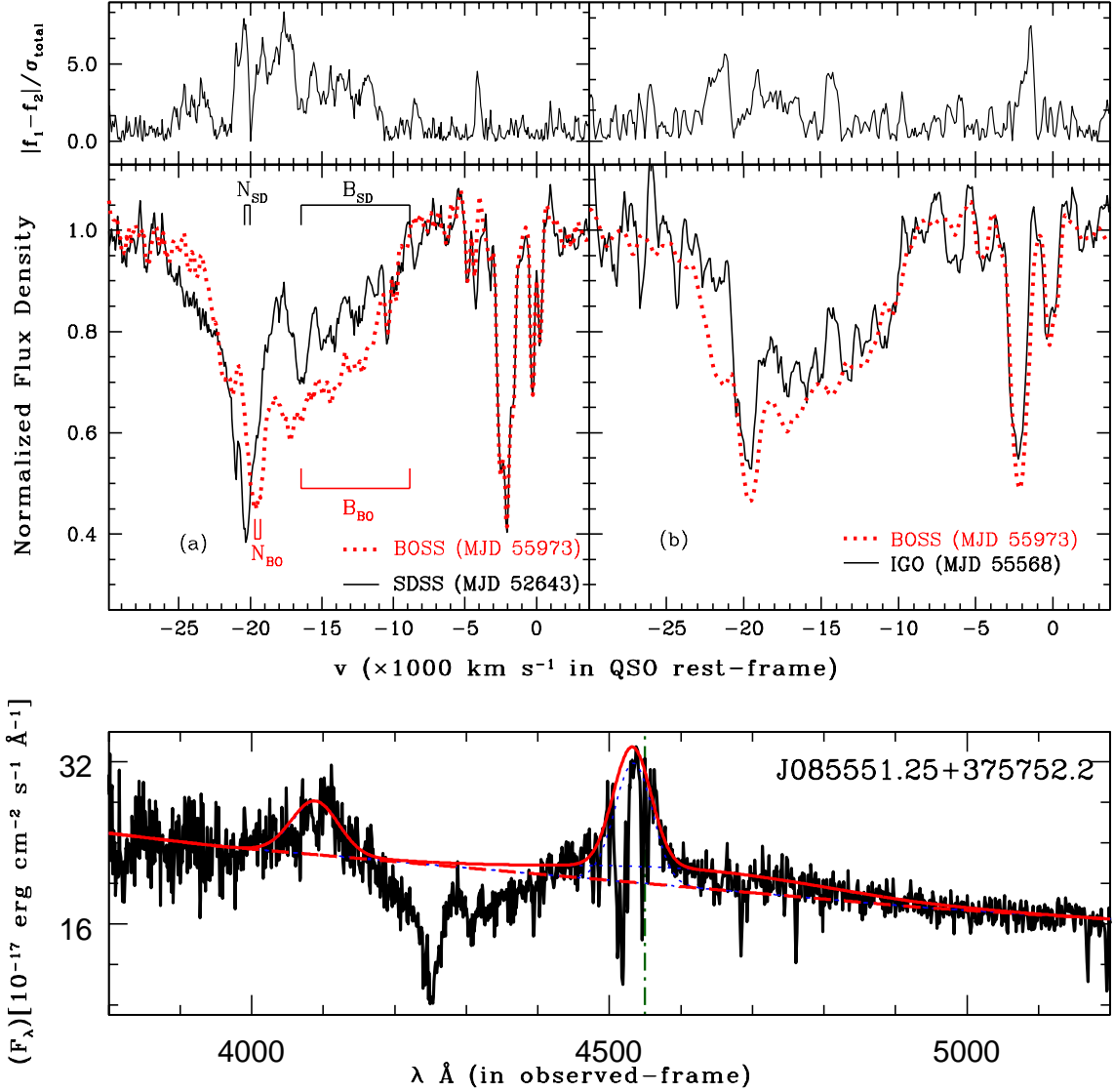,height=16.0cm,width=16.0cm,angle=00,bbllx=55bp,bblly=184bp,bburx=568bp,bbury=691bp,clip=true} 
   \caption{{Lower panel:} The final continuum fit (smooth curve)
     comprising of a power law, a lower order polynomial (dashed
     curve) and the multi-Gaussian components (dotted curve)  for
       SDSS (MJD 52643) spectrum. {Middle panel:} Two-epoch
     absorption line variation in a continuum-normalized SDSS, BOSS
     and IGO spectra for SDSS J085551$+$375752 in velocity scale, with
     $ v =0 $ \kms\ corresponding to QSO emission redshift of \zem\ $=
     1.936$. {Upper panel:} gives the ratio of absolute deviation to
     the total error-bars.}
\label{fig:xbal_vary_0855}
\end{figure*}

\begin{figure*} 
 \centering
 \epsfig{figure=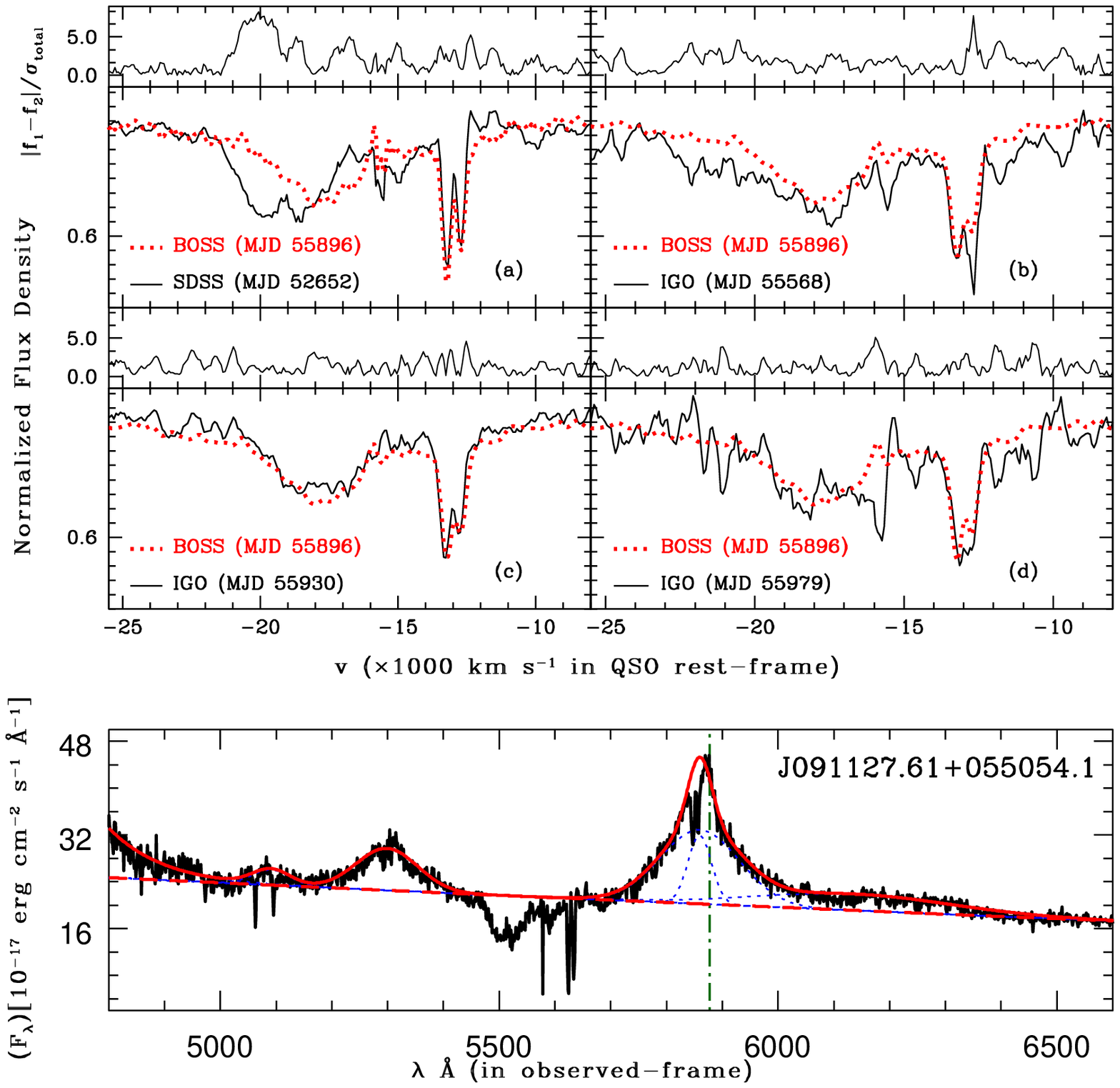 ,height=16.0cm,width=16.0cm,angle=00,bbllx=55bp,bblly=184bp,bburx=568bp,bbury=691bp,clip=true} 
     \caption{ Same as in Figure~\ref{fig:xbal_vary_0855},  for the
         SDSS (MJD 52652) spectrum of SDSS J091127$+$055054, {upper
         panel:} In velocity scales, $ v =0 $ \kms\ correspond to the
       QSO emission redshift of \zem\ $= 2.793$.}
 \label{fig:xbal_vary_0911}
 \end{figure*}

   In order to measure the absorption line variability, it is
  imperative to take into account any emission line flux variation as
  well. For this we also carry out simultaneous fit\footnote{To carry
    out the simultaneous fit we have used the \textsc{mpfit} package
    for nonlinear fitting, written in \textsc{Interactive Data
      Language} routines. \textsc{mpfit} is kindly provided by Craig
    B. Markwardt and is available at
    \url{http://cow.physics.wisc.edu/\~craigm/idl/.}} of emission
  lines with multiple Gaussian profiles without associating any
  physical meaning to them. However, fitting the \civ\ emission lines
  has additional complications due to (i) its asymmetric line profile
  and (ii) the presence of absorption features in the emission line
  region which makes the estimation of QSO continua highly uncertain.
  Therefore, first we masked the wavelength regions having absorption
  signature and then used between one to three Gaussian to define the
  \civ\ line profile over the spectra already normalized with our
  power law fit (see above). All other emission features in the
  spectrum, such as (i) the iron emission blend redward of the
  \civ\ emission line, especially in the range $1500 - 3500$~\AA
  ~\citep[e.g.,][]{VandenBerk2001AJ....122..549V}; (ii) the
  \siv\ emission line and any other feature similar to the emission
  line profile are modelled with a single Gaussian. Lower panels
    in Fig.~\ref{fig:xbal_vary_0855} and ~\ref{fig:xbal_vary_0911}, shows
    our final continuum fit (the dashed line) comprising of a power
    law, a lower order polynomial and the multi-Gaussian components,
    for J085551$+$375752 and J091127$+$055054 respectively. \par

To estimate the spectral variability, the highest signal-to-noise
(S/N) ratio BOSS spectrum, is selected as a reference 
  spectrum for comparing with SDSS and IGO spectra. Both the
SDSS and BOSS data have similar spectral resolution, however, while
comparing two spectra with different resolution such as SDSS/BOSS
spectra (i.e., FWHM $\sim 2.5$~\AA) with IGO (i.e., FWHM $\sim
4.4$~\AA) we have degraded the higher resolution spectra to the lower
one by using an appropriate Gaussian smoothing. In the following
  section we provide detailed analysis of the two systems. \par

\section{Results} 
\label{lab:xbal_RES}

\subsection{J085551$+$375752}
\label{lab:xbal_res_0855}

   The optical spectrum of J085551$+$375752, plotted in
   Fig.~\ref{fig:xbal_vary_0855}, shows a distinct BAL trough of
   \civ\ at \zabs $\sim 1.746$, with a BALnicity index
   \citep[BI,][]{Weymann1991ApJ...373...23W} of 1296
   \kms\ \citep{Streblyanska2010AIPC.1248..513S} and an absorption
   index \citep[AI, ][]{Hall2002ApJS..141..267H} of 1482
   \kms\ \citep[][]{Trump2006ApJS..165....1T}. However, the
   corresponding BAL troughs for other high-ionization species such as
   \siv\ and \nv\ are not seen in our spectrum. Its optical to X-ray
   spectral index, $\alpha_{ox}$, is found to be $- 1.78$ which is
   larger than the typical value of soft X-ray weak QSOs having
   $\alpha_{ox} < -2$ \citep{Giustini2008A&A...491..425G}. \par

   The pseudo-continuum normalized spectra of J085551$+$375752 is
   plotted in upper panel of Fig.~\ref{fig:xbal_vary_0855} in velocity
   scale, with $ v =0 $ \kms\ corresponding to reference redshift of,
   \zem\ $= 1.936$. The plots show the comparisons between the SDSS,
   IGO and the reference BOSS spectrum over an elapse time of $\Delta
   t=9.12$, $1.11$ yr in  the observed frame, respectively. For
   visual clarity spectra are smoothed over five pixels. The plot shows
   that the BAL trough has a striking variation over a velocity
     range from $- 29521$ to $- 5670$~\kms, although the change in the
     equivalent width (\ew) is very nominal to be about $18 \pm 1$ per
     cent. The quoted error is only the statistical error resulting
     from the fitting process and does not include any possible
     systematic error such as from the continuum placement
     uncertainties.    \par

    It can be noted from Fig.~\ref{fig:xbal_vary_0855} that the BAL
    trough seen in SDSS (MJD 52643) spectrum is composed of  one
      narrow component at $\sim$ $- 20230$~\kms\ (marked as N$_{\rm
      SD}$)  and one broad component ranging from $-16454$ to
      $-8850$~\kms\ (marked as B$_{\rm SD}$, see
      Fig.~\ref{fig:xbal_vary_0855}). The corresponding components in
    BOSS/IGO spectrum are subscripted with `BO', like `SD', the subscript
    used to refer the SDSS spectrum. The narrow component in the SDSS
    spectrum N$_{\rm SD}$ has shown a  redward kinematic shift of
      $\sim$~675~\kms\ as marked N$_{\rm BO}$ in BOSS spectrum. 
      However, the broad component in the SDSS spectrum (B$_{\rm
        SD}$), has grown in the strength (marked as B$_{\rm BO}$)
      between the SDSS and IGO (MJD 55568) epoch. Furthermore, no
      clear signature of deceleration is seen for the narrow component
      (N$_{\rm SD}$) after the IGO epoch (e.g., see
      Fig.~\ref{fig:xbal_vary_0855}, panel \emph{b}), although a
      change in the overall absorption strength is noticeable between
      the IGO (MJD 55568) and BOSS (55973) epoch. \par

   Also it appear that  the line profile of N$_{\rm SD}$ component in
   the SDSS spectrum is nearly similar to the N$_{\rm BO}$ component in
   the IGO/BOSS spectrum, apart from the above constant velocity
   shift. This shift along with the likely invariance of line-profile of this
   component gives a hint of the outflow deceleration, which we have
   estimated from the position of their line centroid to be $\sim
   - 217.6 \pm 22.2$~\kmsy (i.e., $\sim - 0.7 \pm 0.1$ \cmss), where
   for error, a conservative value corresponding to
   uncertainty of one pixel-shift has been taken.

\subsection{J091127$+$055054}
\label{lab:xbal_res_0911}

 J091127$+$055054 is a gravitationally lensed QSO at \zem\ $= 2.793$
 \citep{Bade1997A&A...317L..13B}. It is also an X-ray bright radio
 quiet QSO having X-ray luminosity of $4 \times 10^{46}$\erg
 \citep{Bade1997A&A...317L..13B} and $\alpha_{ox}$ of $- 1.58$
 \citep{Giustini2008A&A...491..425G}. Chandra  observations of this
 source on 1999-11-02 and 2000-10-29, show that its 0.5 $-$ 8.0 keV
 flux  is invariant over the rest-frame time span of about $95$ days
 \citep{Gibson2012ApJ...746...54G}. Its optical spectrum shows a
 distinct BAL trough of \civ\ at \zabs\ $\sim 2.549$, with a BI value
 of $2358$ \kms\ \citep{Streblyanska2010AIPC.1248..513S} and an AI of
 $1149$ \kms\ \citep{Trump2006ApJS..165....1T}. \par

\begin{figure}
\epsfig{figure=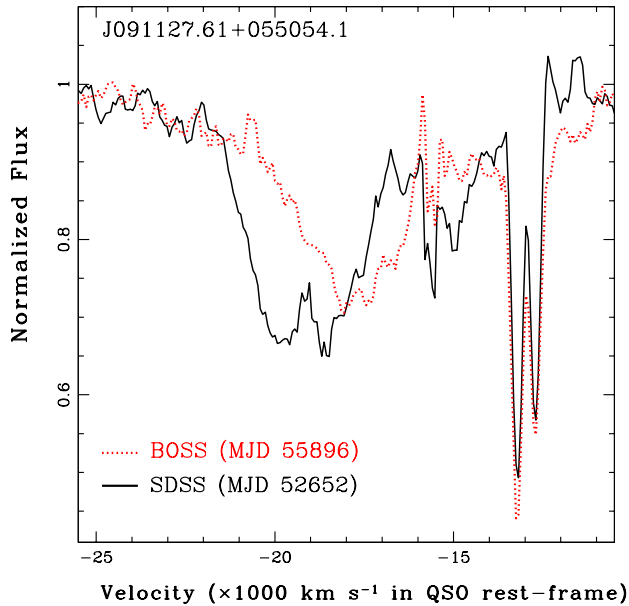,height=8.5cm,width=8.5cm,angle=0}
 \caption{The radial velocity shift observed in J091127.6$+$055054 for
   \civ\ line between SDSS (MJD 52652) (solid thick curve)
   and BOSS (MJD 55896) epoch (dotted curve) spectrum.}
\label{fig:J091127p05505_oplot}
\end{figure}

\begin{figure}
\epsfig{figure=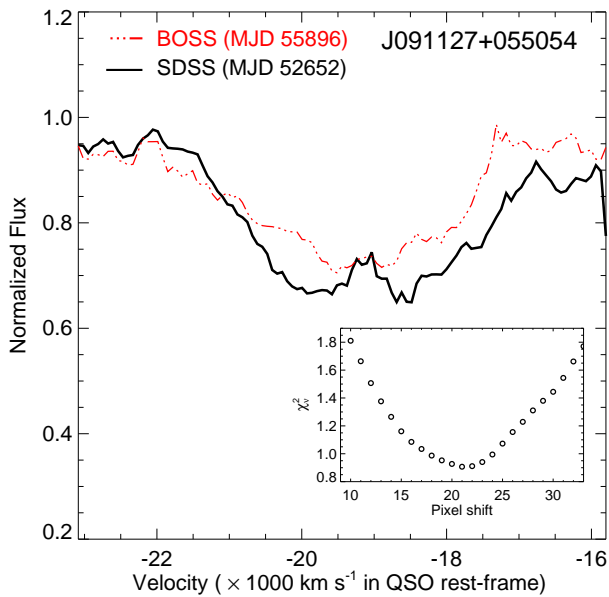,height=8.5cm,width=8.5cm,angle=0}
 \caption{Overplot of SDSS and BOSS epoch spectra for BALQSO
   J091127.6$+$055054 after applying best-fitting pixel shift of 21
   pixels on BAL trough of BOSS spectrum; the inset displays the
   \chinu\ versus pixel shift curve.}
\label{fig:pixshift}
\end{figure}

 A comparison between SDSS (MJD $52652$), IGO (MJD $55568$, $55930$
 and $55979$) and the reference BOSS (MJD $55896$) spectrum of
 \civ\ BAL trough (all smoothed over five pixels), over  an
   observed time span of about $8.89$ and $0.89$ yr respectively are
 shown in Fig.~\ref{fig:xbal_vary_0911} in velocity scale with $ v =0
 $ \kms\ corresponding to reference redshift of \zem\ $= 2.793$. 
   From this figure the change and shift in line profile is clearly
   evident between the SDSS (MJD $52652$) and the BOSS (MJD $55896$)
   epoch.  Assuming this as a constant velocity shift of the whole
   profile (e.g., see Fig.~\ref{fig:J091127p05505_oplot}, where we
   reproduce the \civ\ BAL trough spectral portion from
   Fig.~\ref{fig:xbal_vary_0911}), one can compute such average shift
   by using cross-correlation technique by minimizing
   \chisq\ \citep[e.g.  see,][and inset in
     Fig.~\ref{fig:pixshift}]{Hall2007ApJ...665..174H}.  Using this
   technique a best-fitting average shift of $21 \pm 1$ pixel (i.e.,
   $\sim 5$~\AA\ in QSO rest-frame) has been found between SDSS and
   BOSS spectrum.  It corresponds to a deceleration of $a = - 2.0 \pm
   0.1$~\cmss, over a rest-frame time span of $2.34$ year. However, no
   such kinematic shift is noticeable between the BOSS and subsequent
   IGO spectra (e.g., see Fig.~\ref{fig:xbal_vary_0911}).  \par

\section{Discussion}
\label{lab:xbal_DnC}

 In view of the fact that BAL trough variability can be a powerful
 tool to probe the physical conditions in the vicinity of the QSO,
 there have been many systematic efforts to detect the \ew\ and the
 absorption centroid variations \citep[][and references
   therein]{Ak2013ApJ...777..168F}. However, until now very few cases
 of variability in the absorption kinematics of BAL troughs have been
 reported \citep[e.g.,][]{Vilkoviskij2001MNRAS.321....4V,
   Rupke2002ApJ...570..588R, Gabel2003ApJ...595..120G,
   Hall2007ApJ...665..174H}. A first detection of a change in radial
 velocity in an outflow associated with a narrow absorption system in
 a Seyfert 1 galaxy, NGC 3783, is discovered by
 \citet{Gabel2003ApJ...595..120G}, where they found a radial shift in
 \civ, \nv, and \siv\ lines. In addition, they also determined a
 deceleration values of $a = -0.25 \pm 0.05$ and $-0.10 \pm
 0.03$~\cmss\ over the rest-frame time span of 0.75 and 1.1 yr
 respectively. Similarly, \citet{Hall2007ApJ...665..174H} have
 reported the largest BAL trough acceleration in SDSS
 J024221.87$+$004912.6 at $a = 0.154 \pm 0.025$ ~\cmss over a
 rest-frame time span of 1.39 yr.\par

Our investigation is related to the BAL variability of less explored
X-ray bright BALQSOs, where we have monitored 10 well selected
  high-ionization X-ray bright BALQSOs (see
  Section~\ref{lab:xbal_intro}). Here, we have reported  a
    discovery of such rarely seen kinematic shift in the \civ\ BAL
    trough for two members of our sample, namely J085551$+$375752 and
    J091127$+$055054.

As pointed out in Section~\ref{lab:xbal_intro}, some of the main
mechanisms proposed to understand such absorption strength and/or
kinematic shift in absorption profiles are (i) by actual line-of-sight
acceleration/deceleration of a shell of material from a continual
flow; (ii) changes in the geometry of the outflow,  such as
  directional shift of outflow; and, (iii) changes in the ionization
state as a function of velocity in a fixed outflow. These
  mechanisms of BAL trough variations are briefly discussed below, as
  an attempt to constrain the nature of the outflow in our two X-ray
  bright BALQSOs.

\subsection{Deceleration of a continual flow}
\label{lab:subsec_Decel_var}

One possible explanation for the observed decrease in radial velocity
is that the absorbing cloud is undergoing a bulk radial deceleration.
We first consider gravity as the source of the required inward radial
force. The simplest case that the radial force has decreased relative
to gravity between the two intervals, can be tested first by
estimating the distance of absorbing cloud from the central mass and
then carrying out the comparison between outflow speed and the escape
velocity at the absorbing cloud location. With an average outflow
speed of $\sim$ 20,000 \kms, absorbing cloud of J085551$+$375752 and
J091127$+$055054, would have moved by a distance of $\sim$ $2.1 \times
10^{17}$ cm and $\sim$ $1.5 \times 10^{17}$ cm, in a rest-frame time
span of 3.11, 2.34 yr, respectively, between SDSS and IGO/BOSS epoch.
In addition, studies based on photoionization modelling suggests that
BAL cloud may be at a distance of several pc to Kpc from the central
ionizing source \citep[e.g.,
  see][]{Bautista2010ApJ...713...25B,Rozanska2014NewA...28...70R}.
Hence, assuming a conservative limit of $\sim$ 1 pc for the distance of
absorbing cloud from the central ionizing source at the SDSS epoch,
the total distance at IGO/BOSS epoch (after including the distance
traveled by an outflow since SDSS epoch as estimated above) would be
$\sim$ $3.3 \times 10^{18}$ cm and $\sim$ $3.2 \times 10^{18}$ cm, for
J085551$+$375752 and J091127$+$055054 respectively. Given that the
central black-hole mass for J085551$+$375752 and J091127$+$055054 is
about $6.03 \times10^9 M_{\odot}$ and $8.31 \times10^9 M_{\odot}$,
respectively \citep{shen2011ApJS..194...45S}, we found that the
respective escape velocities at the distance of their \civ\ BAL will
be around $\sim$ 6970, $\sim$ 8261 \kms. Therefore, being both these
escape velocities much smaller than the typical observed outflow speed
of $\sim 20,000$~\kms, exclude the possibility of gravitational force
being the cause of the deceleration of the outflow and may likely be
related to the non-gravitational forces (e.g., see below). \par

\subsection{Directional shift in the outflow}
 \label{lab:subsec_Driec_var}

  Another simple possibility of the observed decrease in the radial
  velocity is that the absorbing cloud is moving along a curved path
  across our line of sight and we are looking at the changing radial
  component of the velocity vector along our line of sight
  \citep[e.g.,][see their figure 3]{Gabel2003ApJ...595..120G}. Such
  curved path trajectories for the BAL outflows are commonly predicted
  by various mechanisms, such as, the absorber is driven off the accretion
  disc by the radiation pressure from the central source
  \citep[][]{Murray1995ApJ...451..498M, Proga2000ApJ...543..686P} or
  by the magneto-hydrodynamic driven winds \citep[e.g.
    see,][]{Fukumura2010ApJ...723L.228F,Kazanas2012AstRv...7c..92K}.
  For instance, the magnetically driven disc wind model of
  \citet{Fukumura2010ApJ...723L.228F} predicts the position of
  \civ\ ion in winds, which shifts outward along line of sight of
  decreasing position angle with the disc plane. Further, depending on
  the observer's inclination, the geometric shape of the magnetic field
  lines and ionization equilibria may result in the substantial
  observable change in width/shift of absorption trough,
  provided clouds motions are curved enough to account for the observed 
  velocity changes (in range of $\sim$ 217 $-$ 620 \kmsy).
   For an observed outflow speed of $\sim 20,000$~\kms, the predicted
   inclination angle by \citet{Fukumura2010ApJ...723L.228F} is less
   than 50 degrees and hence for most of the duration while moving
   along the sight line, absorbing cloud will be covering the background UV
   continuum source. This avoids the possibility of otherwise changing
   optical depth also due to probable variation of partial-coverage.

	 \subsection{Photoionization driven BAL variation}
	 \label{lab:subsec_Phto_var}

         In this scenario, the underlying hypothesis is that the
         unsaturated BAL trough could vary in response to the changing
         ionization state of QSO central engine, which leads to
         estimation of the minimum electron density and maximum
         distance from the continuum source for a photoionized plasma
         \citep[e.g.,][]{Hamann1997ApJ...478...87H,Hamann1998ApJ...500..798H,Narayanan2004ApJ...601..715N}.
         \par

For J085551$+$375752, assuming the observed variability time-scale
(i.e., 3.11 yr in the QSO rest-frame) as an estimate for recombination
time-scale ($\rm t_{rec}$), then the photoionization equilibrium would
imply the electron number density, n$\rm _ e$, to be $ \ga 3000$
cm$^{-3}$(using $\rm n_e \ga [1/(\alpha_r t_{rec})]_{C~{\sc
    IV}}$).  Further, the presence of the \civ\ ions implies that the
ionization parameter, $U$, should be such that log$~U > - 2$
\citep[e.g., see][]{Hamann1997ApJ...478...87H}.  This in conjunction
with the fact that log$~U \propto \rm 1/n_e r^2 $ have allowed us to
estimate the distance, $r$, of the absorbing cloud from the QSO centre
to be $\la$ 3 Kpc.

In addition, no significant continuum variation is found, while
comparing SDSS and IGO spectrum (which has better flux calibration
than BOSS spectrum). On the other hand, absence of \siv\ absorption
suggest that the ionization parameter range should be such that when
the radiation field increases, the \ew\ of \civ\ (or its column
density) should decrease (e.g., see figure 1 of
~\citealt{Hamann1997ApJ...478...87H}) at IGO/BOSS epoch, which indeed
is the case as at IGO/BOSS epoch the \civ\ trough got weakened. So, perhaps, 
the variability at optical region might be very mild (within
uncertainty) but the variability in UV region (that ionizes \civ) may be
appreciable, though, more observational constrains will be helpful to
conclude firmly about this possibility. \par

Similarly, for outflow seen in our second source J091127.6$+$055054,
assuming the rest-frame variability time-scale (i.e., 2.34 yr here),
as an estimate for recombination time-scale, we have computed the
electron number density and the distance of an absorbing cloud from
the center of J091127.6$+$055054 to be $\ga 4000$ cm$^{-3}$ and $\la$
5 Kpc, respectively (using the approach as discussed above for the
case of J085551$+$375752). The continuum flux for this source at
$5000$~\AA\ has decreased by a factor of about 1.7 between the SDSS
and the IGO epoch spectra. Such decrease in the ionizing radiation is
also hinted by the \civ\ emission line that appears stronger (i.e.,
higher \ew\ due to decrease in continuum) in the IGO/BOSS spectrum
than that of earlier epoch SDSS spectrum. We also note here, in
Fig.~\ref{fig:pixshift}, that the overall velocity spread of
\civ\ absorption remains same (apart from shift) between the two
epochs, however, the overall absorption strength has weakened. The
absorption from other species such as \siv\ and \nv\ associated to the
\civ\ BAL trough are not detected in our spectrum. As a result one
would expect the allowed ionization parameter range to be such that
when radiation field decreases (i.e., dimming of QSO continuum) the
absorption line strength of \civ\ should increase \citep[e.g.,
  see][]{Hamann1997ApJ...478...87H}. In contrast to this, our
\civ\ trough got weakened (e.g., see
Fig.~\ref{fig:J091127p05505_oplot}) with the dimming of our continuum
source at IGO/BOSS epoch as compare to SDSS epoch; suggesting that
photoionization driven variation mechanism to be unlikely for this
case. \par

 In addition to the above discussed scenarios, there are also fair chances
 for other mechanisms at play as well. For instance, spray of small
 clouds causing their appearance/disappearance or evolution due to
 cloud irradiation \citep[e.g., see][]{Proga2014ApJ...780...51P},
 along with range of intrinsic velocities, may also result in the
 observed kinematic shift and strength variability in BAL trough,
 causing change in resultant optical depth even without variation of
 flow-speed over time.

\section{Conclusions}
\label{lab:xbal_con}

In this paper, we present the spectra of two BALQSOs, namely
J085551$+$375752 and J091127$+$055054, whose broad \civ\ absorption
lines have shown a shift in velocity, between observation spanning a
rest-frame time of 3.11 and 2.34 yr, respectively. The shift in the
velocity is generally very rare in BALQSOs, and on top of that both
these QSOs belong to a rare sub-class of X-ray bright BALQSOs. The
X-ray irradiation of any absorbing cloud might play a crucial role in
their evolution. For instance, at high X-ray luminosity ($\rm L_X $)
limits, \citet[][]{Barai2012MNRAS.424..728B} shows that inner gas will
significantly get heat up and expand, which can result into a strong
enough outflow capable of expelling most of the gas at larger radii.
On the other hand, for some intermediate $\rm L_X$, thermal
instability might even induce a non-spherical feature with cold and
dense clumps surrounded by over heated clouds. Similarly,
\citet{Fan2009ApJ...690.1006F} also found that the BAL properties such
as BI and maximum outflow velocities do correlate with the intrinsic
X-ray weakness, and suggests that X-ray absorption might probably be
necessary for observed BAL outflows.

\par

For our X-ray bright BALQSOs J085551$+$375752 and J091127$+$055054,
the radial shift we found in our spectrum, resulted in outflow
deceleration of about $\rm a = - 0.7 \pm 0.1, -2.0 \pm 0.1$~\cmss,
which is about a factor of 2.8 and 7.8 larger than the highest
deceleration value reported till now by
\citet{Gabel2003ApJ...595..120G} for NGC 3783 with $\rm a = - 0.25 \pm
0.05$~\cmss. We also explore here the main possible mechanisms such as
(i) deceleration of a continual flow; (ii) directional shift in the
outflow and (iii) the photoionization induced BAL variation, along
with other possibility such as, spray of small clouds along the line
of sight. We found that the deceleration due to continual flow to be
an unlikely mechanism based on our estimation of absorbing clouds
distances from the central mass, where the escape velocity is too
small in comparison to the outflow speed (see
Section~\ref{lab:subsec_Decel_var}). The photoionization induced BAL
variation may be a possibility for J085551$+$375752, but seems
unlikely for the variation seen in J091127$+$055054 (e.g., see
Section~\ref{lab:subsec_Phto_var}). On the other hand, the directional
shift in the outflow can equally be an explanation for the observed
radial shift in both cases, provided the motion is curved enough for
observed large velocity shift (about $\sim$ 217 $-$ 620 \kmsy). In
view of the fact that both of our BALQSOs are X-ray bright, probably
their BAL troughs may favor the small clouds scenario, rather than a
single homogeneous continuous radial outflow, in conjunction with the
scenario of directional shift of outflow motion. However, many small
clouds also allow the possibility of other mechanisms such as spray of
small clouds causing their appearance/disappearance, with or without
curved path motion, can also result in the observed kinematic and/or
strength variability in BAL trough, due to the change in optical depth
even without variation of speed over time. For the scenario with
curved path motion, small clouds close to the disc might contribute,
but for scenario with spray of clouds without curve path motion the
clouds at higher inclination might be more effective, and hence
resultant of them might be responsible for the observed outflow
variation seen in our two sources. \par

We also note that, generally, the effects of covering factor,
ionization and optical depth are complexly related, so disentangling
these effects would further require the variability follow-ups along
with the high-resolution observation as well as a dense sampling of
time-variability over larger wavelength coverage, in addition to our
present observational constrains. Beside this, as pointed earlier that
the majority of line variation do not exhibit velocity shift and these
two rare cases which further belong to rare sub-population of X-ray
bright BALQSOs, make it difficult to interpret their outflow
variation in terms of the global physical properties of BAL outflows.
Nonetheless, given the importance of such observed kinematic shift to
constrain the possible QSO outflow models, we conclude that
J085551$+$375752 and J091127$+$055054 are potential candidates to
investigate outflow by monitoring them both in optical as well as in
X-rays for their lines and/or continuum variability. Also, a systematic
search, in large spectroscopic surveys like SDSS/BOSS, for BALQSOs
having such kinematic shift and BAL strength variation, will further
help to increase the sample of such cases of varying outflows and
improve our general understanding of QSO outflows.

\section*{Acknowledgments}

   We thank an anonymous referee for his/her constructive report
    to improve our manuscript. \par

We gratefully acknowledge the observing help rendered by Dr. Vijay
Mohan and the observing staff at the IGO 2-m telescope. JM wishes to
thank the Indian Academy of Science for their support through Summer
Research Fellowship and grateful for hospitality at ARIES. \par

   Funding for the SDSS and SDSS-II has been provided by the Alfred P.
   Sloan  Foundation,  the  Participating Institutions,  the  National
   Science  Foundation, the  U.S. Department  of Energy,  the National
   Aeronautics and Space  Administration, the Japanese Monbukagakusho,
   the Max  Planck Society, and  the Higher Education  Funding Council
   for England. The SDSS Web Site is http://www.sdss.org/. The SDSS is
   managed   by  the   Astrophysical  Research   Consortium   for  the
   Participating Institutions. The  Participating Institutions are the
   American  Museum   of  Natural  History,   Astrophysical  Institute
   Potsdam, University of Basel, University of Cambridge, Case Western
   Reserve  University,  University  of  Chicago,  Drexel  University,
   Fermilab, the Institute for Advanced Study, the Japan Participation
   Group, Johns  Hopkins University,  the Joint Institute  for Nuclear
   Astrophysics,  the Kavli  Institute for  Particle  Astrophysics and
   Cosmology,  the  Korean Scientist  Group,  the  Chinese Academy  of
   Sciences   (LAMOST),   Los    Alamos   National   Laboratory,   the
   Max-Planck-Institute for Astronomy (MPIA), the Max-Planck-Institute
   for  Astrophysics (MPA),  New Mexico  State University,  Ohio State
   University,  University of  Pittsburgh,  University of  Portsmouth,
   Princeton University, the United  States Naval Observatory, and the
   University of Washington.

\bibliography{references}

\label{lastpage}
\end{document}